# A Frugal Model for Accurate Early Student Failure Prediction

Ikram Gagaoua

*Domoscio by RiseUp, Université de Lorraine, CNRS, LORIA*
Paris, France
ikram.gagaoua@riseup.ai

Armelle Brun

*Université de Lorraine*
*CNRS, LORIA*
Nancy, France
armelle.brun@loria.fr

Anne Boyer

*Université de Lorraine*
*CNRS, LORIA*
Nancy, France
anne.boyer@loria.fr

## Abstract

*Predicting student success or failure is vital for timely interventions and personalized support. Early failure prediction is particularly crucial, yet limited data availability in the early stages poses challenges, one of the possible solutions is to make use of additional data from other contexts, however, this might lead to overconsumption with no guarantee of better results. To address this, we propose the Frugal Early Prediction (FEP) model, a new hybrid model that selectively incorporates additional data, promoting data frugality and efficient resource utilization. Experiments conducted on a public dataset from a VLE demonstrate FEP's effectiveness in reducing data usage, a primary goal of this research. Experiments showcase a remarkable 27% reduction in data consumption, compared to a systematic use of additional data, aligning with our commitment to data frugality and offering substantial benefits to educational institutions seeking efficient data consumption. Additionally, FEP also excels in enhancing prediction accuracy. Compared to traditional approaches, FEP achieves an average accuracy gain of 7.3%. This not only highlights the practicality and efficiency of FEP but also its superiority in performance, while respecting resource constraints, providing beneficial findings for educational institutions seeking data frugality.*

## 1. Introduction

Predicting student success or failure is a critical task. Early prediction not only highlights risk factors but also empowers educators and educational institutions to proactively identify students at a high risk of academic failure. This proactive approach extends beyond mere detection, enabling the deployment of timely and specifically tailored interventions. These strategic measures are vital in preventing potential academic challenges and promoting an environment of equitable educational opportunities for all students [6] [3]. However, such an early prediction raises a significant challenge related to the limited availability of data which negatively impacts the accuracy of the predictions [1]. To address this issue, we make the following hypothesis, in line with [9]: incorporating additional data from other sources, during the early stages, can significantly enhance prediction accuracy. Such other sources can be for example other courses the student attended, previous achievements, demographic data, etc.

Yet, it is essential to acknowledge that incorporating additional data comes at a cost, not only computational and financial but also environmental. In this context, we aim to design the first education-related prediction model that limits the amount of data used, especially regarding additional data collected. Concretely, the model is intended to be frugal. Data frugality can be defined as a strategic approach that centers on minimizing resource consumption, both in terms of data and computational resources, while still achieving acceptable or optimal performance [7]. It groups various strategies, including optimizing data acquisition and utilization, making learning processes more efficient, and reducing the memory and computational demands of machine learning models. In our primary focus on data frugality, we aim to minimize data consumption while optimizing predictive accuracy.

This work is therefore driven by the following research question:

**(RQ) How to design a frugal model that accurately early predicts student success or failure?**

To pursue such a frugal approach of early predictive modeling, we introduce the Frugal Early Prediction (FEP) model, a new hybrid model designed to manage several data sources, while conditionally using them. The condition on which the data selection relies is the estimated predictive accuracy of the model. FEP is thus set apart from traditional methods [5] [9] [6] that uniformly and systematically employ one or several data sources for every prediction. By adopting this condition-based approach, we aim to optimize the use of relevant data, i.e. minimizing data consumption. To the best of our knowledge, FEP is the first model that considers frugality in the frame of education. In the following sections, we describe the FEP model, the methodology, and the evaluation, including the dataset selection, and data pre-processing.

Experiments are also presented, confirming frugality and highlighting an unexpectedly increased performance of the FEP model in terms of accuracy and stability.

## 2. Related Work

The field of Artificial Intelligence in education has witnessed significant advances in learning outcome prediction. Various studies have explored the prediction of student success or failure using different approaches and techniques. In 2019, Abyaa et al. conducted a systematic review of student modeling literature over the past five years [2]. Their review focuses on student learning and performance prediction. They emphasize the significance of understanding student characteristics and incorporating them into predictive models for accurate performance prediction. Building upon this, Wang et al. conducted a study that exploits data from a virtual learning environment (VLE) to predict university students at risk of academic challenges [11]. The research uses students' clickstream data and assessment results as features to develop a predictive model using the well-known Open University Learning Analytics dataset "OULAD" [8]. The findings demonstrate the effectiveness of machine learning algorithms in identifying students at risk. More importantly, the study highlights the issue of limited data availability in the early stages of the course, obviously in terms of achievements, but also in terms of clickstream data. This limitation is in line with the issues considered in our work; to address it, the literature identifies three distinct approaches. This section will explore each of these methods, providing a comprehensive understanding of how they contribute to tackling data limitations. Abu Zohair et al. acknowledge the importance of data by exploring the prediction of student performance using small dataset sizes, emphasizing the challenges associated with limited data availability [1]. The study highlights the importance of leveraging advanced modeling techniques to overcome the limitations posed by small datasets.

Volariç et al., on the other hand, explore the use of data augmentation with Generative Adversarial Networks (GANs) to address the issue of data scarcity. They aim to improve at-risk students' prediction in a VLE [10]. The study showcased the potential of GANs in generating synthetic data to overcome small datasets and enhance the accuracy of prediction models. We acknowledge the effectiveness of data augmentation techniques, particularly in their application to various fields. However, we also recognize the importance of carefully assessing their suitability for specific contexts, such as education and human learning. The tasks in this domain are unique, often requiring a balance between computational efficiency and the clarity of results. While the advanced capabilities of data augmentation methods bring undeniable benefits, their computational intensity and complexity might pose challenges in educational settings. These settings often prioritize resource efficiency, ease of understanding, and transparent methodologies. Therefore, in environments where simplicity, cost-effectiveness, and clear explication are essential, alternative approaches that meet these criteria may offer significant value. To address the issue of data scarcity, leveraging additional data sources has shown promise. Costa et al. demonstrate this through their study on the effectiveness of educational data mining techniques in the early prediction of academic failure in introductory programming courses. By utilizing a diverse range of data sources, including student demographics and learning interactions, their research highlights the potential of these techniques in identifying at-risk students and enabling timely interventions. This approach emphasizes the value of incorporating various data types to enhance prediction accuracy [6]. However, it is crucial to consider the associated costs and practical challenges related to the collection, retrieval, and training of additional data. Samuelsen et al. highlight the benefits of integrating data from different sources, such as learning management systems (LMS), online forums, and assessment records [9]. This extensive data collection likely results in a substantial volume of data. In contexts where data consumption is a priority, such a model might face challenges in terms of adaptability and efficiency, due to its extensive data requirements. To conclude, these works collectively underscore the significance of accurate prediction models in educational settings and highlight various approaches and methodologies employed in the field. While they provide key findings, they often face challenges related to limited data availability in early prediction and adopt various solutions to cope with this limit, however, many of these solutions have the disadvantage of consuming a lot of data, which poses challenges related to frugality and even privacy.

## 3. FEP: a Frugal Early Prediction Model

We aim to address the challenge of early prediction with limited data and propose the FEP (Frugal Early Prediction) model that leverages additional data to improve prediction accuracy, while optimizing data consumption. To the best of our knowledge, the challenge of minimizing data consumption while ensuring effective predictive modeling has not been addressed in the field of education. The originality of the FEP model lies in the student-level data selection strategy. Recall that in the literature, a source-level strategy is generally adopted (a source is either completely used or not used at all). The key concept of FEP revolves around using the right data at the right moment, specifically tailored to meet the needs and profiles of individual students. FEP represents a novel contribution to the field, by offering a customized data management approach.

FEP is characterized by a personalized selection of data sources. Concretely, the selection (and use) of data from a data source is conditioned to the student-level estimated predictive accuracy of the model and is intended to enhance the accuracy of predictions. A key aspect of the FEP model is that once data is retrieved from an additional source, it can be utilized throughout the entire prediction process, without additional cost. This approach is applicable through various algorithms such as XGBoost, SVM, Decision Trees, etc. FEP is inspired by two models from the literature: a single-source model and a multi-source model. These models, detailed below, contextualize our contribution.

### 3.1 The Single Source (SS) Model

The single source model (SS), which will be considered as a baseline model, exclusively utilizes a unique primary data source for the prediction task. This model reflects the predictive capability achievable solely with the available data from the selected context (e.g. a course). This model is a supervised model that relies on a single data source. The workflow of this model is represented in Figure 1 (a).

### 3.2 The Multi-Source (MS) Model

The Multi-Source (MS) model exploits systematically, for all students, all data available, possibly from several data sources. The MS model relies on the hypothesis that the more data available, the more accurate the model. MS is a supervised model that uses all the data available for all the learners. The MS model is represented in Figure 1 (b).

### 3.3 The Frugal Early Prediction (FEP) Model

The Frugal Early Prediction (FEP) model, that we introduce in this work, is a semi supervised two-phase model. The first phase performs prediction by using a single data source only, the primary source, in a supervised approach. This prediction is associated with a student-level estimation of the accuracy of this prediction. The model used in this phase is the SS model. The estimation of the accuracy of this prediction is based on a heuristic that is dataset dependent. The semi-supervised characteristic of the model comes from the use of this heuristic. A low estimated accuracy value for a given a student means that the system is not confident in its prediction. In this case, a second phase is run. In this second phase, additional student-level data sources are integrated and used to predict student success or failure.

The prediction model used in this second phase can be viewed as a Multi-Source (MS) model. By using additional data, the prediction accuracy is expected to be increased.

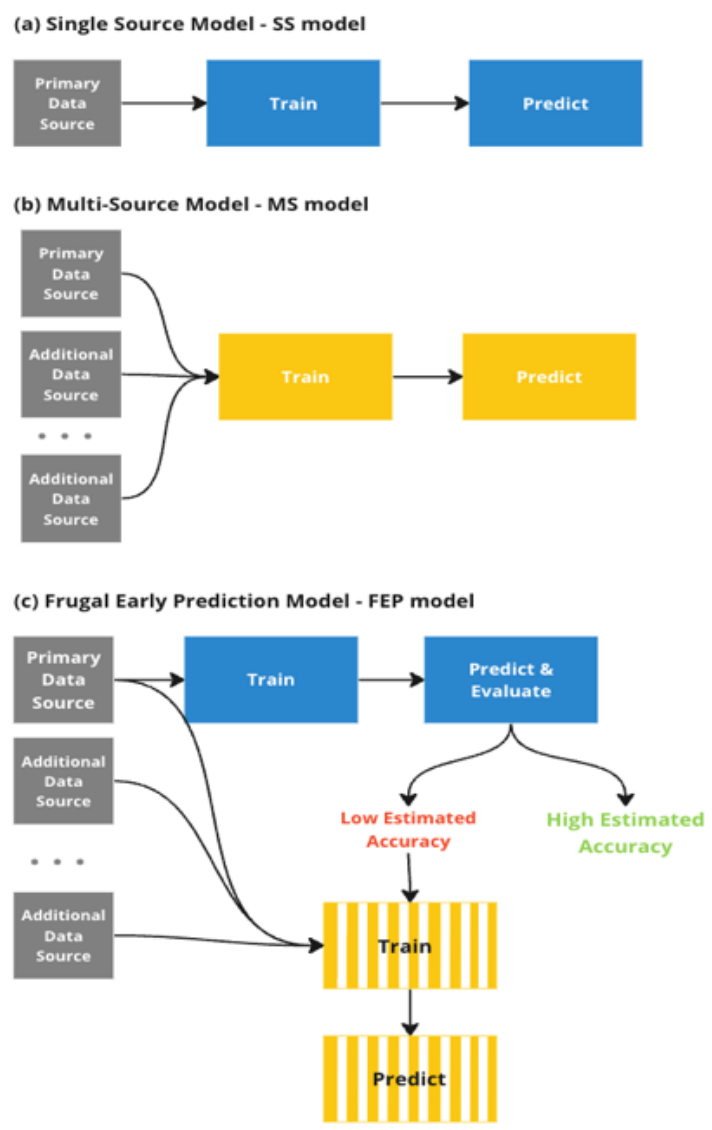

**Figure 1.** Workflow of the models

The conditional integration of additional data sources is central to the FEP model's hybrid nature. The structure and workflow of the FEP model are presented in Figure 1 (c). Figure 2 presents three examples of prediction scenarios with FEP. Associated data usage and consumption are also displayed. This figure highlights the condition under which additional data is retrieved and when the data is re-used later. In the case of student 1, only the primary data source is used on the 1st evaluation. The associated estimated accuracy of the prediction (in this case, a success) is high. The system thus does not exploit additional data. The same holds for the following evaluations. In the case of student 2, on the 3rd evaluation, the estimated accuracy is low. Thus, the model gets and exploits additional data for student 2. Once this additional data is integrated, the estimated accuracy is high enough. In this case, the use of external data has contributed to increase the estimated accuracy.

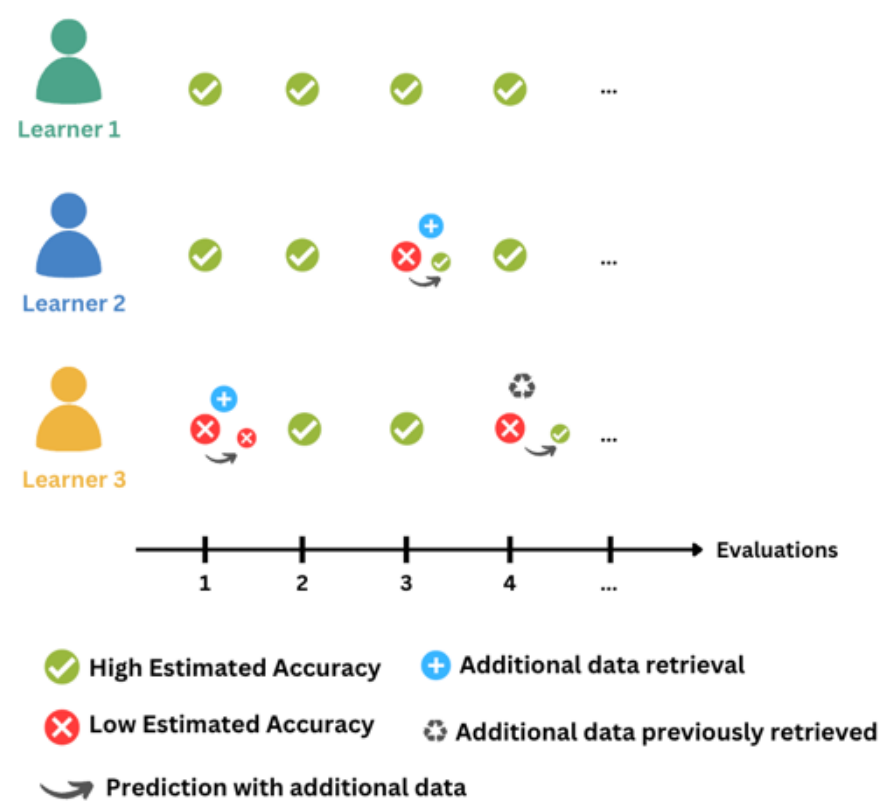

**Figure. 2.** FEP Model data consumption

Yet, student 3 has multiple instances where the estimated accuracy is low. For example, additional data retrieval is performed for the 1st evaluation. However, the estimated accuracy remains low. On the 4th evaluation, the same additional data is re-used and contributes to a high estimated accuracy.

# 4. Evaluation Metrics

We will adopt a multi-dimensional evaluation of FEP, that encompasses accuracy, temporal aspects, and data frugality.

### 4.1. Accuracy

Accuracy is an essential metric used to evaluate the performance of models in machine learning. It quantifies the proportion of total predictions that a model correctly classifies. This metric is crucial in various domains, as it provides a straightforward measure of how well a model is performing in terms of correctly identifying or predicting outcomes.

### 4.2 Earliness Stability Score (ESS)

The Earliness Stability Score (ESS) is a time-related metric introduced by [3], to evaluate a model within educational settings. It is designed to capture two critical aspects of prediction quality: how early the system can make correct predictions (earliness) and how consistently it can make these correct predictions over time (stability). ESS combines these two dimensions into a single score that helps in assessing the trade-off between being early and being consistent in predictions.

**Earliness** measures the earliest point in time at which a predictive model can make a correct prediction. It is crucial in contexts where early predictions lead to more effective outcomes. It is calculated as the time point of the first correct prediction made by the system. Earliness is evaluated by the following equation:

$$Earliness = \frac{1}{n}\sum_{i=1}^{n} e_i \qquad (2)$$

Where:
– n is the total number of students.
– $e_i$ is the first time point where prediction is correct for student i.

**Stability** measures the consistency and reliability of a model's predictions over time. It assesses the model's robustness in maintaining accurate predictions throughout the course. It is defined as the average length of the longest sequence of consecutive correct predictions for each student

$$Stability = \frac{\sum_{i=1}^{n}|h(s_i)|}{|n|} \qquad (3)$$

Where:
– n is the total number of students.
– $|h(s_i)|$ is the length of the sequence of correct predictions for a given student i.

**ESS** is calculated as the harmonic mean between the average earliness and average stability. ESS is computed as:

$$ESS = \frac{2\times(1-earliness)\times stability}{(1-earliness)+stability} \qquad (4)$$

ESS provides meaningful insights into the model's performance, considering both the timeliness of predictions (earliness) and the reliability of its results (stability). By considering these crucial aspects, the ESS score provides a comprehensive assessment of the model's effectiveness in educational settings. By definition, a high ESS value reflects a high ability to consistently deliver early accurate predictions over an extended duration.

### 4.3 Frugality: Data Consumption

The data consumption metric is defined as the total amount of data parts used per n students. In the literature, while metrics address the use of data in general predictive modeling scenarios, there is a lack of specific metrics that account for multi-source data usage. This metric seeks to bridge that gap by quantifying data consumption in various modeling contexts, particularly emphasizing the distinction in data usage when integrating multiple data sources versus a single source, and is defined as follows:

$$DataConsumption = \left(\frac{\sum_{i=1}^{n} Q_i}{n}\right) \qquad (5)$$

Where:
– n is the total number of students.
– $Q_i$ the number of data source used for each student i.

This metric is crucial for understanding and comparing the data utilization of different models using additional sources in terms of their data usage strategies

# 5. Experimentation

### 5.1 Dataset Selection and Description

The OULAD dataset. is popular in the education domain. It is associated with the Open University's online learning platform, a Virtual Learning Environment (VLE) [8]. It groups data from seven selected courses (or modules) offered by the university.

The dataset includes information about student demographics, such as location, age group, disability, education level, and gender. It also provides insights into student interactions with the VLE, including accessing course content, participating in forum discussions, submitting assessments, and checking assignment marks. Additionally, the dataset contains student assessment marks, which reflect their performance in assignments or exams. Overall, the OULAD dataset offers valuable information for analyzing student engagement, behavior, and academic achievement within the Open University's online courses.

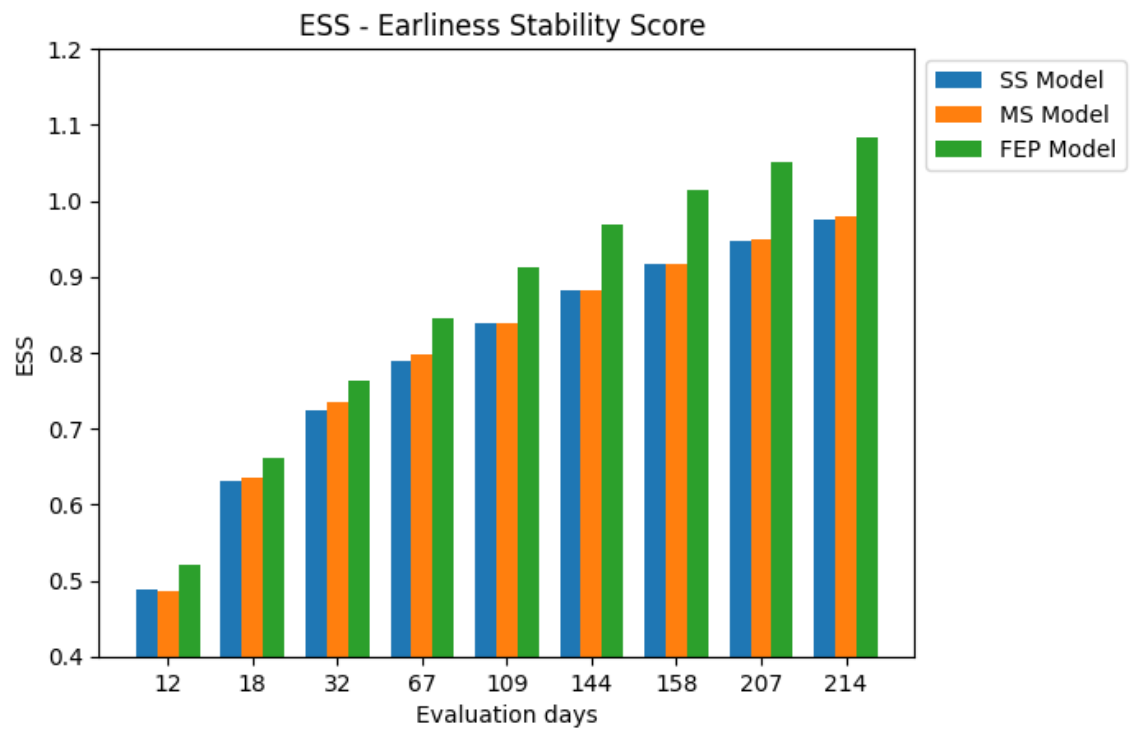

**Figure 3**. ESS results

### 5.2. Data Pre-processing

The objective of the pre-processing we perform is to prepare the data for analysis. Concretely, we select the appropriate semester and course and model relevant attributes to enhance the prediction process.

**Selection of primary and additional data sources**

Considering the dataset chosen for our work, we strategically choose to focus on the spring semester of 2014 (2014J), which is the last semester in the dataset. This decision is guided by the need to have a comprehensive dataset encompassing a substantial number of students with a prior course history. We use the data related to a specific course from starting that semester as the primary data source. Specifically, we focus on the CCC course (2014J-CCC), which lasts 269 days, as it has the highest number of students who have taken another course previously, making it an ideal choice for analyzing their performance and behavior. The additional data source is made up of data from a course from the students' historical academic records, in our case, the previous course completed by the student in the past. The resulting primary source data is composed of 694 students (378 succeeded & 316 failed).

### 5.3. Classification Model

To predict the success or failure of each student in the next assignment, we explored various classification algorithms, including SVM, Decision Trees, and Random Forest, to evaluate their efficacy in the specific prediction task of this work. Through comparative evaluations, it became apparent that the XGBoost algorithm [4] stood out as the most effective in handling tabular data and generating highly accurate predictions. Its superior performance and robustness made it the optimal choice for our prediction in this study.

## 6. Results & Discussion

In the experiments, we run and compare the SS, MS, and FEP models on the result prediction task. We present here the associated results, findings, and outcomes. We analyze and discuss the performance of these models based on the selected evaluation metrics, providing key observations into their effectiveness and practical applicability.

### 6.1 Accuracy

The prediction accuracy of the three models is shown in Table 1. First, we can see that accuracy does not show a progressive increase over time, for any model. This means that using a longer time frame to predict the success or failure of the next evaluation does not yield a significant positive impact on accuracy. Second, contrary to our expectation, the accuracy of the multi-source (MS) model is not consistently higher than that of the single-source (SS) model. For example, in the middle of the course duration, i.e. on the 67th and 109th days, the SS model performs better than the MS model (accuracy is 0.94 and 0.92 respectively). This finding suggests that the systematic use of additional data is not fruitful for every student or in every prediction context.

As expected, the frugal model (FEP) consistently demonstrates an enhanced accuracy over the SS model. More importantly, FEP achieves a systematic improvement over the MS model. This improvement is on average of 7.3%.

Table 1. Accuracy of the SS, MS and FEP models

| Periods | 7th | 12th | 18th | 32nd | 67th | 109th | 144th | 158th | 207th | 214th |
|---|---|---|---|---|---|---|---|---|---|---|
| SS Model | 0.85 | 0.87 | 0.87 | 0.83 | **0.94** | 0.89 | 0.88 | 0.88 | 0.86 | 0.87 |
| MS Model | 0.86 | 0.86 | 0.89 | 0.83 | **0.92** | 0.87 | 0.90 | 0.89 | 0.89 | 0.86 |
| FEP Model | 0.94 | 0.96 | 0.94 | 0.93 | **0.98** | 0.94 | 0.94 | 0.93 | 0.92 | 0.93 |

### 6.2 Earliness Stability Score - ESS

The ESS (Earliness Stability Score) for the SS, MS and FEP models are displayed in Table 2 for different time frames throughout the course, and Figure 3 is a graphical representation of the table.

Table 2. ESS of SS, MS, and FEP models

| Periods (day) | 12th | 18th | 32nd | 67th | 109th | 144th | 158th | 207th | 214th |
|---|---|---|---|---|---|---|---|---|---|
| SS Model | **0.49** | 0.63 | 0.72 | 0.79 | 0.84 | **0.88** | 0.92 | 0.95 | 0.98 |
| MS Model | **0.49** | 0.64 | 0.74 | 0.80 | 0.84 | **0.88** | 0.92 | 0.95 | 0.98 |
| FEP Model | **0.52** | 0.66 | 0.77 | 0.85 | 0.91 | **0.97** | 1.01 | 1.05 | 1.08 |

FEP outperforms baseline models (SS and MS) for the ESS metrics, throughout all the time frames. On the 12th day, both the SS and MS models achieve the same ESS (0.49), while the FEP model achieves a slightly higher ESS with 0.52 .This suggests that at very early points, the targeted additional data slightly improves the model. As the course progresses, ESS increases for the three models and the FEP model consistently and increasingly outperforms the other models (11% improvement in the last time frame).

FEP's higher ESS score suggests its superior ability to identify critical trends and indicators in the educational data well earlier than baseline models. This early detection capability is crucial in educational settings, as it allows for timely interventions that can significantly alter a student's learning trajectory. Additionally, the stability implied by this high score indicates that the FEP model's predictions are consistent and reliable over time, even as the course content and student behavior evolve. Such stability is essential in ensuring that the predictive results remain relevant and actionable throughout the course duration.

### 6.3. Data Consumption

Let us recall that the MS model systematically uses additional data for prediction, whereas the SS model never uses additional data and relies only on the primary source. The FEP model adopts a conditional approach for student-level additional data usage. In the dataset used, 323 students (from 694 students total in the dataset) have at least one inaccurate prediction among different time frames. Additional data is thus collected and used for these 323 students only. Additional data is thus incorporated for only 46% of students. According to the data consumption metric, the single-source model, consumes n data points for n students, equating to a 1:1 data-to-student ratio. In contrast, the multi-source model utilizes 2n data, indicating a doubled data consumption rate per student. Meanwhile, the FEP model, a more efficient approach, consumes 1.46n data points for n students, demonstrating a moderated increase in data usage compared to the single source but a decrease compared to the multisource, as the FEP model consumes 27% less data than the multi-source. This selective retrieval allows the FEP model to focus on a specific subset of students with inaccurate predictions, resulting in more accurate predictions.

## 7. Conclusion

In conclusion, our work confirms the pertinence of a selective use, in comparison with a systematic use, of additional data. On an educational-related dataset, the proposed FEP model uses a significantly reduced amount of data (reduced by 27%) and thus contributes to the frugality objective. In addition, we have shown that the prediction accuracy is improved (by 7.3% on average), which confirms that more data is not always better, at least for student success or failure prediction. Importantly, the earliness stability of the model is higher than the one of the two other models studied (around 10% improvement), confirming that a selective use of additional data makes prediction models more stable in earlier time frames.